\documentclass[smallextended]{svjour3}       
\smartqed  
\usepackage{graphicx}
\usepackage{multirow}
\usepackage{alltt}

\usepackage{amsmath} 
\DeclareMathOperator*{\argmax}{arg\,max}
\usepackage{amssymb,amsfonts}
\usepackage{colortbl}
\usepackage{pifont}
\usepackage{graphicx}
\usepackage{textcomp}
\usepackage{booktabs}
\usepackage{algorithm}
\usepackage{algpseudocode}
\usepackage{fancyvrb}
\usepackage{multirow}
\usepackage[table,xcdraw]{xcolor}
\usepackage[utf8]{inputenc}
\usepackage{fancyhdr}
\usepackage{url}
\usepackage{adjustbox}

\usepackage{wrapfig} 
\usepackage{changepage} 
\usepackage{framed}
\usepackage{enumitem}
\usepackage{cite}

\usepackage{graphicx}
\usepackage{longtable}

\definecolor{niceblue}{HTML}{0000FF}

\newcommand{\BLUE}{\color{niceblue}}
\newcommand{\BLACK}{\color{black}}

 \usepackage[hidelinks]{hyperref}

\newcommand{\there}[1]{%
  \hyperlink{resp:#1}{%
    \fcolorbox{black}{black!15}{%
      \bfseries\scriptsize #1%
    }~on page \pageref{resp:#1}%
  }%
}

\newcommand{\here}[1]{%
  \hypertarget{resp:#1}{}
  \fcolorbox{black}{black!15}{%
    \label{resp:#1}%
    \bfseries\scriptsize  {#1}%
  }%
}

\definecolor{lightblue}{rgb}{0.85,0.9,1}
\definecolor{blue}{rgb}{0.5,0.75,1}
\definecolor{darkblue}{rgb}{0,0.5,1}

\definecolor{lightred}{rgb}{1,0.85,0.85}
\definecolor{red}{rgb}{1,0.5,0.5}
\definecolor{darkred}{rgb}{1,0,0}

\definecolor{codeblue}{rgb}{0.1,0.1,0.7}
\definecolor{codegreen}{rgb}{0.1,0.6,0.1}
\definecolor{codegray}{rgb}{0.5,0.5,0.5}
\definecolor{codepurple}{rgb}{0.58,0,0.82}
\definecolor{backcolour}{rgb}{0.95,0.95,0.92}
\definecolor{verylightgray}{rgb}{0.95, 0.95, 0.95}

\usepackage{bm}
\renewcommand{\BLUE}{\BLACK}
\renewcommand{\there}[1]{}
\renewcommand{\here}[1]{}

\begin{document}
\newcommand{\bi}{\begin{itemize}}
\newcommand{\ei}{\end{itemize}}
\title{From Brittle to Robust: \\Improving LLM Annotations for SE Optimization 
}


\author{Lohith Senthilkumar         \and
        Tim Menzies 
}


\institute{All authors are from Computer Science, North Carolina State University
   Oval Dr,  Raleigh, NC 27606
\\\email{lohithsowmiyan1@gmail.com, timm@ieee.org}}

\date{Received: date / Accepted: date}

\journalname{Empirical Software Engineering}
\maketitle

\begin{abstract}
Software analytics often builds 
from labeled data.
Labeling   can be slow, error prone, and expensive.
When human expertise is scarce,
SE researchers sometimes
ask large language models (LLMs) for the   missing labels.

While this has been successful in some domains,
recent results show that LLM-based  labeling has blind spots.
Specifically, their labeling is not effective for  higher dimensional multi-objective problems.

To address this task, we propose a novel LLM prompting strategy called SynthCore. When one opinion fails,  SynthCore's   combines multiple separated opinions generated by LLMs (with no knowledge of each others' answers) into
an ensemble of few-shot learners. 
Simpler than other strategies (e.g. chain-of-thought, multi-agent-debate, etc)   SynthCore   aggregates results from    multiple single prompt sessions (with no crossover between them). 

  SynthCore has been tested on 49 SE multi-objective optimization tasks,
 handling tasks as diverse as software project management, Makefile configuration, and hyperparameter optimization.
SynthCore's
ensemble   found optimizations
that are better than state-of-the-art alternative approaches  (Gaussian Process Models, Tree of Parzen Estimators,
active learners in both exploration and exploitation mode). 
Importantly, these optimizations were made using data labeled by LLMs, without any human opinions.

From these experiments, we conclude that ensembles of few shot learners can successfully annotate high dimensional multi-objective tasks. Further, we speculate that other successful few-shot prompting results could be quickly and easily enhanced  using SynthCore's ensemble approach.

To support open science, all our data and scripts are available  at \url{https://github.com/lohithsowmiyan/lazy-llm/tree/clusters}.

\end{abstract}

\section{Introduction}

Software Engineering (SE) data is often costly and labor-intensive to obtain, typically requiring careful curation by subject matter experts (SMEs) to ensure both accuracy and relevance~\cite{lustosa2024learning}. Yet, as demand grows for large-scale, high-quality datasets across a range of SE tasks, researchers have increasingly turned to automation to reduce annotation costs~\cite{ahmed2024can}. Among these, the use of Large Language Models (LLMs) as automated oracles for labeling has emerged as a compelling option~\cite{wan2024tnt}.

However, LLM-based annotation is not without challenges. In a recent empirical study presented at the  MSR 2025
conference, Ahmed et al.~\cite{ahmed2024can} systematically evaluated LLM performance on five SE labeling tasks. Their findings reveal a nuanced picture: while LLMs can at times match human-level agreement on straightforward labeling tasks, they frequently fall short on higher dimensional or ambiguous problems. Key issues include hallucinated outputs, unjustified confidence in incorrect answers, and brittleness when applied to unfamiliar or out-of-distribution inputs. Crucially, Ahmed et al. warn that even when LLMs exhibit high inter-annotator or human-model agreement, this is not a reliable proxy for ground-truth correctness-particularly in domains that demand deep software-specific reasoning. They argue that LLMs should be used as conditional collaborators rather than drop-in replacements for human annotators: 

\begin{quote}
{\em ``We do not claim that LLMs can universally replace human annotators; instead, our findings suggest that they may be viable complements—especially when \underline{\bf used carefully} and \underline{\bf selectively.}.}~\cite{ahmed2024can}
\end{quote}
\BLUE
\here{R3c} This paper reports a {\em careful} and {\em selective} use of LLMs for annotating examples used in {\em multi-objective optimization} tasks. By multi-objective optimization, we refer to problems where the goal is not to optimize a single output but to simultaneously balance several (often competing) objectives. This differs from the tasks studied by Ahmed et al., where the output space is typically single-dimensional. In contrast, our optimization setting requires annotations that guide trade-offs across multiple domain constraints. For instance, when evaluating software design options, practitioners may need to find alternatives that allow them to:
\bi
\item deliver the {\em most} features,
\item in the {\em least} time,
\item at the {\em lowest} cost,
\item with the {\em fewest} defects.
\ei
Other examples of multi-objective problems in SE include:
\begin{itemize}
\item
{\em Green engineering} ensures faster responses that use less energy;
\item
and {\em Hyperparameter optimization} finds learner parameters that minimize false alarms and maximize recall;
\item
And any 
{\em Configuration task}
such as  selecting
magic control variables within a Makefile.
\end{itemize}
Modern software urgently
needs better tools for automatic configuration. We say this since all to often, software is deployed with suboptimal configurations~\cite{Krishna:2016}.  For example, \textsc{Storm}'s defaults yield $480\times$ performance degradation vs. optimal parameters~\cite{DBLP:conf/mascots/JamshidiC16}. Such poor performance is hardly surprising since industrial optimization faces major obstacles:
\begin{itemize}
    \item Configuration spaces explode exponentially (in \textsc{7z}: 14 parameters = one million configurations).
    \item Performance landscapes are rugged and sparse~\cite{DBLP:journals/tosem/GongC25,lustosa2024learning,chen2026promisetune}, creating local optimal traps.
    \end{itemize}
    Evaluating a wide range of different configurations
    can be very costly; e.g. \textsc{x264}'s 11 parameters need $1,536$ hours to explore~\cite{DBLP:conf/wosp/ValovPGFC17}, limiting budgets to dozens of evaluations~\cite{DBLP:journals/tse/Nair0MSA20,DBLP:conf/icse/0003XC021}.
 {\em Active learning} can reduce that cost by evaluating only the most informative examples. But the success of any optimizer—including active learning—can depend heavily on the availability of reasonable initial examples. In active learning, the learner selectively queries or acquires labels, but its ability to do so effectively often hinges on a high-quality initial set of annotated samples. These initial samples, commonly referred to as ``warm starts,'' serve as the starting point from which the learner begins exploring the space of candidate solutions.
 
Generating such warm starts has proven to be challenging. Recent work~\cite{senthilkumar2024can} shows that LLMs often struggle to produce high-quality initial examples, particularly in domains that are unfamiliar or underrepresented in their training corpora. We have previously argued that this limitation arises from insufficient training data coverage:
\bi
\item LLMs excel when tasks are frequent, well-represented, and structurally consistent with patterns in their training data.
\item But when confronted with rare, highly specialized, or high-dimensional optimization problems, their outputs tend to be shallow, generic, or systematically misinformed.
\item For details of these findings, see 
Table~\ref{lo0},~\ref{med0}~\ref{hi0}.
\ei
\BLACK
To overcome this limitation, this paper proposes a novel prompting strategy called SynthCore.
Rather than relying on a single monolithic answer from the LLM, SynthCore prompts the model multiple times—each time under slightly different contexts or with randomized seeds—and then synthesizes the resulting outputs into a composite candidate set. 
 Simpler than other strategies (e.g.
chain-of-thought, multi-agent debate, etc) \cite{yuan2025ui2html,yang2024chain,du2024multi} SynthCore aggregates results from multiple
single prompt sessions (with no crossover between them).\BLUE \here{R2a-1} It also differs substantially from self-consistency prompting strategies that rely on a voting based mechanism to select the majority output while decoding. \cite{ahmed2023better}. Moreover, prior work on seed generation \cite{xia2024fuzz4all} has shown that varying few-shot examples encourages language models to produce more diverse seeds—an insight that we regard as central to our method. \BLACK By aggregating diverse perspectives from the LLM, SynthCore avoids the pitfalls of overconfident single-shot reasoning and improves the chances of capturing useful problem structure.

We evaluated SynthCore on 49 multi-objective SE optimization tasks spanning domains such as project planning, Makefile tuning, and hyperparameter selection. Across the board, SynthCore’s ensemble-based strategy delivered superior performance compared to state-of-the-art techniques, including Gaussian Process Models, Trees of Parzen Estimators, and both exploitative and exploratory active learners. Notably, all labels used in SynthCore’s experiments were generated solely by LLMs—no human annotations were included. These results demonstrate that,
\begin{quote}
{\em
While individual LLM responses may be unreliable, {\bf a carefully curated ensemble of few-shot learners} can collectively overcome their shortcomings.}
\end{quote}
Thus, we conclude that with the right orchestration, LLMs can move from brittle annotators to robust contributors in SE automation pipelines.

To introduce SynthCore, this paper explores two research questions

\BLUE
\here{R3e}
\begin{itemize}

    \item{\textbf{RQ1:} {\em While standard few-shot learning often struggles in SE Active Learning, does ensemble-based few-shot learning overcome these limitations and achieve better performance?} This is the core question of this research.}
    \item {\textbf{RQ2:} {\em Does the effectiveness of ensemble-based few-shot learners hold when dealing with high-dimensional SE data?} This work was motivated by a prior study that reported
    LLMs failed   on
 higher dimensions.} Therefore,
    in our results, we must pay particular
    attention to this kind of data.
    
    \end{itemize}
As seen in the results of this paper: 
\bi
\item
{\bf RQ1} Yes, the results demonstrates the superiority of ensemble LLM Learners  in SE multi-objective optimization tasks.
\item

{\bf RQ2} Yes, the results shows that ensembles of few shot learners solve the problem reported in prior work.  

\ei

 \BLACK
The contributions in this paper include:
\begin{enumerate}
    \item We propose an novel prompting tactic: ensembles of LLM few-shot learners.
\item We test  the effectiveness of this approach for 49 SE datasets.
\item We compare this approach with  alternative methods for 49 datasets.
\item We offer  a reproduction package with all our data and scripts\footnote{  \url{https://github.com/lohithsowmiyan/lazy-llm/tree/clusters}}.
\end{enumerate}

Based on this work,
we speculate
that other successful few-short prompting results could be quickly and easily
enhanced using SynthCore’s ensemble approach.
This is not a proven contribution but the results of this paper
suggest it would be a useful 
future research direction for prompt engineering and software analytics.

  \BLUE 
 
  \subsection{Digression}
  \here{R0a} Before beginning,
 it is important to clearly define the scope of this work. This paper does not investigate which specific LLM architecture is optimal for these optimization tasks. While we have preliminary hypotheses on that matter, a comprehensive exploration of model-specific performance remains a critical direction for future research.
 
 What this paper offers is an important finding regarding  LLMs and optimization: contrary to recent pessimistic results~\cite{senthilkumar2024can}, LLMs are highly effective for optimization problems, even with high-dimensional data. Our results demonstrate the definitive superiority of LLMs over long-standing, state-of-the-art symbolic methods for this task. This is an important contribution to the Software Engineering (SE) literature, which currently suffers from insufficient comparative evaluation of LLMs. For instance, a recent systematic review of 229 SE papers using LLMs found that only $13/229 \approx 5\%$ compared LLMs to other approaches~\cite{Hou24}. 
 It is important to explore this gap since the prevailing hype often obscures the fact that LLMs frequently do not outperform more established methods in many domains~\cite{grinsztajn2022why,somvanshi2024survey,Tawosi23,majumder2018500+,Fu17,johnson2024ai}. What we can offer here  is good news since we can offer  robust evidence that LLMs offer a significant, practical performance advantage over conventional methods in the domain of SE optimization.

\section{Literature Review}
\subsection{Annotations in Software Engineering}
Manual annotations are vital for data-driven Software Engineering (SE)
tasks, influencing empirical findings \cite{easterbrook2008selecting}. Annotated
datasets underpin research in defect prediction, vulnerability detection,
sentiment analysis, and static analysis validation
\cite{menzies2025retrospective}. Redundancy, using multiple annotators,
ensures reliability and mitigates bias \cite{morse2002verification}.
Examples include 11 annotators for Jupyter Notebook inconsistency
\cite{patra2022nalin} and 3 annotators (from 11) for Java method
similarity \cite{kamp2019sesame}. Manual labeling is costly. Generic
automated tools often fail in SE due to specialized terminology
\cite{7962370}, leading to the development of domain-specific models like SENTI-STRENGTH
(78\% accuracy, 85\% recall) \cite{7962370}.

Annotation quality is crucial but faces challenges:
\begin{itemize}
\item \textbf{Heuristic Labeling Pitfalls}: Lexical patterns (e.g., "bug",
"fix") for defect prediction yield inconsistent ground truth without
rigorous validation \cite{vasilescu2015quality}.
\item \textbf{Validation Discrepancies}: Manual errors are common; 90\% of
false positives in one technical debt analysis were labeling errors
\cite{9226105}.
\end{itemize}
Manual annotation can be very slow.
\begin{itemize}
\item 
Unlike
medical labs where automated agents   label  based
on clearly defined definitions~\cite{krasowski2014autoverification}, SE annotation is far more subjective.  
\item
Valerdi reports studies with panels of human experts
in software effort estimation. Those panels
could handle only 13.3
examples/hour \cite{valerdi2010heuristics}.
\item
Lustosa reports studies with humans were complex management ranking
peaked at only 15-20 judgments/hour \cite{lustosa2024learning}. 
\item
In prior work with industrial subject matter experts,
we found expert availability
was very limited (rarely $>3$ hours/week) \cite{menzies1992expert}. Also, when we could negotiate
for more access, those longer
sessions decreased label quality \cite{EASTERBYSMITH19803}
(due to cognitive fatigue).
\end{itemize}
\subsection{Can LLMs Replace Manual Annotators in SE?}
In theory, automated annotation methods address manual limitations
\cite{dollmann2016and}. Large Language Models (LLMs) are promising
due to their pre-training on vast corpora, which allows them to adapt to
domain-specific contexts without extensive local training. LLMs are
used in code generation \cite{svyatkovskiy2020intellicode} and testing
\cite{deng2023large}.

Ahmed et al. \cite{ahmed2024can} explored if LLMs can replace manual
annotators on 10 SE tasks (e.g., code summarization, semantic similarity,
static analysis labeling). LLMs achieved \textbf{moderate to substantial
agreement} with humans in \textbf{6 out of 10 tasks} (e.g., non-functional
requirements) \cite{ahmed2024can}. However, performance declined
significantly in context-intensive tasks (e.g., static analysis warnings),
showing low model-model agreement, suggesting LLMs cannot fully
replace humans, especially for nuanced tasks.

Senthilkumar et al. \cite{senthilkumar2024can} reached similar
conclusions about the limits of LLMs for labeling.
That study
used LLM annotations to generate \textit{warm-start} samples for
\textit{active learning} in multi-objective SE tasks.
They found LLMs   useful for low-dimensional probems, but not
high-dimensional, multi-objective optimization problems
(this conclusion persisted even after applying feature
synthesis to artifically  reduce dimensionality). 


Senthilkumar et al. used the MOOT repository (\cite{Moot:2025}, see Table~\ref{dataset}), a collection
of SE multi-objective optimization tasks with $x$ independent variables
(3 to 38) and $y$ goals (1 to 5). Datasets were categorized by
$x$-dimensionality: \textit{Low} ($|x| < 6$), \textit{Medium} ($6 \le |x| \le 11$),
and \textit{High} ($|x| > 11$) \cite{difiore2024}. They found:

\begin{itemize}
\item For \textbf{Low-dimensional tasks}: LLM warm-starts matched or
surpassed symbolic traditional  (Gaussian Process and TPE methods.~\cite{bergstra2011algorithms}).
\item For \textbf{Medium-dimensional problems}: same result;
\item For \textbf{High-dimensional problems}: LLM warm-starts underperformed,
often falling below baselines \cite{senthilkumar2024can}.
\end{itemize}
They concluded that  LLMs struggle to generalize to complex, high-dimensional
optimization, limiting their utility beyond trivial problems. That result,
which is very negative for LLM-based research,
motivates us to explore alternate methods (hence this paper).

\begin{table}
\caption{Data used in this study.
As per  Di Fiore et al.~\cite{difiore2024},
data is labeled low, medium, or high dimensional
based on the number of $x$ independent goals. 
 $|x| < 6$ means "low" ;
 $6 \le |x| \le 11$ means "medium" and  $|x| > 11$  means "high". For more details on this data, see \S\ref{data}.}\label{dataset}
 
\begin{center}
{\footnotesize
\begin{tabular}{p{1.7cm}|lrrl}  
\textbf{Groups} & \textbf{File Name} & $|$rows$|$ & \textbf{$|x|$ / $|y|$} & \textbf{Dimsionality} \\ \hline
\multirow{32}{*}{Configuration} 
    & SS-A    & 864    & 3/2  & low      \\
    & SS-B    & 972    & 3/2  & low      \\
    & SS-C    & 1080   & 3/2  & low      \\
    & SS-D    & 2736   & 3/2  & low      \\
    & SS-E    & 3008   & 3/2  & low      \\
    & SS-F    & 3839   & 3/2  & low      \\
    & SS-G    & 5184   & 3/2  & low      \\
    & SS-H    & 4608   & 4/2  & low      \\
    & SS-I    & 6840   & 5/2  & low      \\
    & SS-J    & 65536  & 6/2  & medium   \\
    & SS-K    & 86058  & 6/2  & medium   \\
    & SS-L    & 1023   & 11/2 & low      \\
    & SS-M    & 864    & 17/3 & high     \\
    & SS-N & 86058 & 18/2 & high \\
    & SS-O    & 972    & 11/2 & high     \\
    & SS-P    & 1023   & 11/2 & high     \\
    & SS-Q    & 2736   & 13/3 & high     \\
    & SS-R    & 3008   & 14/2 & high     \\
    & SS-S    & 3840   & 6/2  & medium   \\
    & SS-T    & 5184   & 12/2 & high     \\
    & SS-U    & 4608   & 21/2 & high     \\
    & SS-V    & 6840   & 16/2 & high     \\
    & SS-W    & 65536  & 16/2 & high     \\
    & SS-X    & 86058  & 11/3 & high     \\
    & Apache AllMeasurements & 192 & 9/1 & medium \\
    & SQL AllMeasurements    & 4653 & 38/1 & high \\
    & X264 AllMeasurements   & 1152 & 16/1 & high \\
    & rs-6d-c3 obj1          & 3840 & 6/1  & medium \\
    & rs-6d-c3 obj2          & 3840 & 6/1  & medium \\
    & sol-6d-c2-ob j1        & 2866 & 6/1  & medium \\
    & wc-6d-c1-ob j1         & 2880 & 6/1  & medium \\
    & wc+sol-3d-c4-ob j1     & 196  & 3/1  & low \\
    & wc+rs-3d-c4-obj1       & 196 & 3/1 & low\\
    & wc+wc-3d-c4-ob j1      & 196  & 3/1  & low \\
    \hline
\multirow{10}{*}{  Process  } 
    & pom3a          & 500    & 9/3 & medium \\
    & pom3b          & 500    & 9/3 & medium \\
    & pom3c          & 500    & 9/3 & medium \\
    & pom3d          & 500    & 9/3 & medium \\
    & xomo\_flight   & 10000  & 23/4 & high \\
    & xomo\_ground   & 10000  & 23/4 & high \\
    & xomo\_osp      & 10000  & 23/4 & high \\
    & xomo\_osp2     & 10000  & 23/4 & high \\
    & coc1000        & 1000   & 17/5 & high \\
    & nasa93dem      & 93     & 22/4 & high \\
    \hline
Project~Health 
    & healthCloseIsses12mths0001-hard & 10000 & 5/1 & low \\
    & healthCloseIsses12mths0011-easy.csv & 10000 & 5/1 & low \\
    \hline
\multirow{3}{*}{Miscellaneous} 
    & auto93         & 398    & 5/3 & low \\
    & Wine\_quality  & 1599   & 10/2 & medium \\
    & HSMGP num      & 3456   & 14/1 & high  
\end{tabular}}
\end{center}
\end{table}

\subsection{Insight into why LLMs fail}
LLM success depends on training data diversity. In SE, data can be
sparse or biased, leading LLMs to inadequate initial hypotheses
\cite{bender2021dangers}. To compensate, \textit{diversity of inference}
uses techniques like prompt ensembles to explore a broader spectrum of
candidate solutions during inference \cite{wang2023selfConsistency}.

Given the last paragraph, it seems  appropriate
to explore
\textit{ensemble learning}~\cite{dietterich2000ensemble} where conclusions
are formed from multiple experts, each working
slightly different versions of the data.
Ensemble learning is know to   boost robustness and predictive accuracy,
especially in few-shot, data-scarce settings \cite{dietterich2000ensemble}.
For example,
ensembles (e.g., Random Forest \cite{breiman2001randomForests}) outperform
single models by resisting noise and generalizing better on
high-dimensional data. 


\BLUE

\subsection{Prompting Strategies}
\here{R2a-2}
A wide range of prompting strategies have been explored in recent literature~\cite{ahmed2022few,le2023log,nashid2023retrieval,li2025structured,yang2024chain,ahmed2023better,shinn2023reflexion}. These strategies vary depending on the nature of the problem space, but several have emerged as particularly influential, including \emph{few-shot prompting}, \emph{chain-of-thought prompting}, \emph{self-consistency}, and \emph{self-reflection} approaches. Our own variant of prompting is described in 
\S\ref{prompt}.

Few-shot prompting~\cite{ahmed2022few,le2023log,nashid2023retrieval} is the standard practice of providing a small number of labeled examples (typically $n = 1$ to $4$--$5$) within the prompt. This helps large language models generalize to previously unseen tasks. Many advanced prompting techniques can be viewed as extensions or modifications of the few-shot paradigm.

Chain-of-thought (CoT) prompting~\cite{li2025structured,yang2024chain} extends few-shot prompting by including step-by-step explanations or reasoning traces along with each example. This encourages the model to follow explicit reasoning structures when solving unseen tasks, often improving interpretability and output quality. However, CoT has a notable drawback: if the provided reasoning path is incorrect or biased, the model may confidently follow and amplify flawed logic.

Self-consistency prompting~\cite{ahmed2023better} addresses the limitations of standard CoT. Instead of generating a single explanation--answer pair, the model generates multiple reasoning paths during decoding. These paths are then aggregated, typically via a voting mechanism, to select the most reliable final answer. This reduces sensitivity to any single, potentially incorrect chain of reasoning.

Self-reflection prompting~\cite{shinn2023reflexion} represents a distinct strategy in which the model iteratively refines its responses. In this setting, the model evaluates its previous outputs across a sequence of interactions (question $\rightarrow$ answer $\rightarrow$ critique $\rightarrow$ revised answer), enabling incremental improvement across iterations.

\begin{table}[!t]
\centering
\BLUE
\footnotesize
\begin{tabular}{|p{2cm}|p{4cm}|p{4cm}|}
\hline
\textbf{Feature} & \textbf{Self-Consistency} ~\cite{ahmed2023better}  &
\textbf{SynthCore} (this paper)\\ \hline
\textbf{Core Logic} & \textbf{Consensus}: Truth is the most frequent
answer. & \textbf{Evolution}: Optimal solutions are rare outliers
(mutations). \\ \hline
\textbf{Driver} & Decoding Sampling (Stochastic paths). & Input
Variation (Diverse seeds/contexts). \\ \hline
\textbf{Goal} & Minimize variance (Reliability). & Maximize variance
(Exploration). \\ \hline
\textbf{Selection} & Majority Vote (Mean/Mode). & Sorting/Ranking
(Best of). \\ \hline
\textbf{Analogy} & A jury voting to agree on a verdict. & Biological
mutations driving evolution. \\ \hline
\end{tabular}
\caption{Comparison of prompting strategies.}\label{consistent}
\end{table}

Our proposed \emph{SynthCore} method may appear superficially similar to self-consistency prompting, but as shown in Table~\ref{consistent} there are important distinctions. 
While both methods utilize multiple inference paths, they are
philosophically and mechanically distinct. \textbf{Self-Consistency}
acts as a consensus engine, filtering out noise to find the most
probable ``average'' truth \ ~\cite{ahmed2023better}. In contrast, \textbf{SynthCore}
acts as an evolutionary search engine, actively generating noise
(mutations) to discover superior ``outlier'' solutions.

Also, the two methods take different approaches to variation:
\begin{itemize}
    \item \textbf{Self-Consistency (Consensus):} Views variation as
    uncertainty to be minimized. It assumes correct reasoning paths
converge~\cite{ahmed2023better} . It relies on a voting mechanism to discard
    outliers and select the majority answer.
    \item \textbf{SynthCore (Evolution):} Views variation as a feature
    to be exploited. Drawing on seed generation theory \cite{xia2024fuzz4all}, it
    uses varied few-shot examples to force diverse outputs. These are
    treated as ``strong mutations'' in a search process, designed to
    escape local optima and find better warm-starts for active
    learners.
\end{itemize}
Also there are differences in how the  mechanism,
specifically decoding vs. input context.
This is to say that the methods intervene at different stages of the generation pipeline.
\begin{itemize}
    \item \textbf{Self-Consistency:} Keeps the input constant. It
    generates diversity solely by sampling multiple reasoning paths
    during the decoding stage \cite{wang2023selfConsistency}.
    \item \textbf{SynthCore:}  Varies the input context. It executes
    multiple independent sessions, each with unique randomized seeds
    or shuffled examples ~\cite{ahmed2023better}. This prevents the model from
    fixating on a single context window's logic.
\end{itemize}
Finally, the final answer is generated in a different way:  
\begin{itemize}
    \item \textbf{Self-Consistency:} Aggregates via \textbf{Voting}.
    The goal is to find the mode of the distribution (the most
    reliable answer)~\cite{ahmed2023better}.
    \item \textbf{SynthCore:} Aggregates via \textbf{Synthesis}. The
    goal is to collect all valid samples and sort them by an external
    objective function (e.g., Chebyshev distance) to find the single
    best candidate.
\end{itemize}
While self-consistency focuses on improving the \emph{consistency} of model outputs, our ensemble method emphasizes \emph{diversity}. The goal is to generate a diverse set of high-quality candidate outputs, which can be interpreted as producing strong ``mutations'' in a search process to move toward near-optimal solutions.

\BLACK

\section{Methods}

Drawing inspiration from these established and emerging ensemble methodologies, the rest of this paper investigates the application of ensembles of few-shot prompts for the synthesis of optimal candidate solutions within active learning frameworks. 

This section describes the data, algorithms, performance measures used in this study.

\subsection{Data}\label{data}

\begin{table}
\caption{Introducing the MOOT Repository. Available on line at \url{http://tiny.cc/moot}.}\label{mooteg}

\small
\begin{tabular}{|p{\linewidth}|}\hline
MOOT (Multi-Objective Optimization Testing) is a collection of software engineering datasets drawn from tasks such as process tuning, database configuration, and hyperparameter optimization for, e.g.  defect prediction models.

Table~\ref{mooteg} shows the general
form of MOOT data.  

\begin{center}
\begin{minipage}{3in}
{\scriptsize
\begin{alltt}
  x = independent values          | y = dependent values
  --------------------------------|----------------------
  Spout_wait, Spliters, Counters, | Throughput+, Latency-
     10,         6,        17,    |    23075,    158.68
      8,         6,        17,    |    22887,    172.74
      9,         6,        17,    |    22799,    156.83
      9,         3,        17,    |    22430,    160.14
    ...,       ...,       ...,           ...,    ...
  10000,         1,        10,    |   460.81,    8761.6
  10000,         1,        18,    |   402.53,    8797.5
  10000,         1,        12,    |   365.07,    9098.9
  10000,         1,         1,    |   310.06,    9421
\end{alltt}
}
\end{minipage}
\end{center}

These tabular data sets are  are divided into independent ($x$) inputs and dependent ($y$) goals.
The first row names each column.
Numeric column names start with an uppercase letter.
Goal names ending in + or - are to be maximized or minimized, respectively.
For example, in the above table,  a system configures {\em Spout\_wait, Spliters, Counters} to maximize 
{\em Throughput} and minimize 
{\em Latency}. 

For illustration purposes, rows are sorted from best to worst based on those goals. But note that in our  experiments, rows are randomized and goal values (y) are initially hidden.\\\hline
\end{tabular}
\end{table}

This study explored dozens of software engineering (SE) multi-objective optimization tasks from the MOOT repository~\cite{Moot:2025}
shown in Table~\ref{dataset}
(with more details in 
Table~\ref{mooteg})\footnote{MOOT = {\em Multi-Objective Optimization Tasks}, a collection of recent SE benchmark problems~\cite{Moot:2025}. \url{http://tiny.cc/moot}.}.  
In MOOT, data sets have:  
  \begin{itemize}
    \item 1 to 5 numeric optimization $y$ goals;
    \item 3 to 38 independent $x$ variables;
    \item 90 to 90,000 data rows.
  \end{itemize}
Following a recommendation from Di Fiore et al.~\cite{difiore2024}, the MOOT datasets are categorized based on their input ($x$) dimensionality:
  \begin{itemize}
    \item \textit{Low:} 12 datasets with $|x| < 6$ independent variables;
    \item \textit{Medium:} 14 datasets with $6 \le |x| \le 11$;
    \item \textit{High:} 19 datasets with $|x| > 11$
    \end{itemize}
    (Aside: We acknowledge that “high-dimensional” is defined differently in other domains.
    For example,   text mining may be considered low-dimensional relative to image processing. In defense of our
    categorization, we note that different studies
   report that active learning algorithms exhibit
    very different performance across data sets divided in this way~\cite{difiore2024,senthilkumar2024can}.)

This MOOT data  divides into four groups:
\begin{enumerate}
\item The {\em Config} directory consists of datasets derived from software engineering literature, specifically those labeled with "SS-*" \cite{nair2016accidental}. These datasets comprehensively capture configurations for various tasks, such as video encoding, with primary objectives including query time and run time. Additionally, Config includes datasets related to database configurations, such as Apache\_AllMeasurements.csv, SQL\_ALLMeasurements.csv, and \newline
X264\_AllMeasurements.csv.  

\item The {\em HPO} directory houses datasets from the hyperparameter optimization domain \cite{lustosa2024learning}. These datasets document the outcomes of random forest regression models trained to forecast metrics such as commits, closed issues, and closed pull requests over a 12-month period for open-source projects hosted on GitHub. The Y-values in these datasets represent the error and accuracy of the model under different hyperparameter settings.  

\item The {\em Process} directory contains datasets originating from software process modeling research \cite{green2009understanding, Me07, menzies2009avoid, port2008using}. The "pom*" datasets capture insights into agile development as studied by \cite{boehm2004balancing}. Specifically, the POM3 model represents requirements as a tree of dependencies that emerge dynamically, akin to elements surfacing from water. This model tracks completion rates, idle times, and development effort as teams navigate evolving tasks. Additionally, the "nasdem" dataset provides real-world data, while "osp2" follows the USC Cocomo model format, offering predictions for development effort, defect rates, and risks in waterfall-style software projects \cite{Me07}.  

\item The {\em Misc} directory includes non-software engineering datasets, such as auto93 and WineQuality, which serve as demonstration tools for presenting MOOT to a broader audience.  
\end{enumerate}
We provide   other meta data on the datasets externally.
\footnote{
\BLUE
\here{R4b}, \here{R4c} Details of the parameters for every dataset are provided here: \url{https://github.com/lohithsowmiyan/lazy-llm/blob/clusters/docs/datasets.pdf}
\BLACK
}

\subsection{Performance Measure}
Our active learners returned a row of options.
To measure the effectiveness
of that row, 
  we use the   {\em
Chebyshev Distance}. This is the  maximum distance between any   
$y$ value of a row to its ideal $y$ value in the dataset. 

  \begin{equation}\label{eq:ch}
d_\text{Chebyshev}(y, o) = \max_{i=1, \ldots, n} \left| y_i - l_i \right|
\end{equation}
For this calculation, we normalized the $y$ values for each goal to 0..1 min..max.
For the data set shown in Table~\ref{mooteg}, those ideal $y$ values are associated with maximal {\em Throughput} and minimal {\em Latency}; i.e. their ideal $y$ values are:
\[
\mathit{ideal} \; \{\mathit{Throughput}, \mathit{Latency}\} = \{1, 0\}
\]
We use Chebyshev since:
\begin{itemize}
\item It is also used by other prominent multi-objective algorithms~\cite{zhang2007moea};
\item It is a ``cruel critique'' that punishes   failure for any optimization goal;
\item We have run all our results with other evaluation measures (e.g. average distance of row goals to the ideal) and our main result
(that SynthCore works for higher dimensional data) still persists.
\end{itemize}

\subsection{ Active Learners}
  
Given   $n$ initially labeled examples, active learning proceeds as follows~\cite{brochu2010tutorial,lustosa2024learning}.:

\begin{enumerate}
\item Acquire the next most informative example based on the current model.
\item Label this example and add it to the training set.
\item Update the predictive model.
\item Repeat until the labeling budget ($B$) is exhausted.
\item Return the \textbf{best} example (where ``best'' is defined as per Equation~\ref{eq:ch}).
\end{enumerate}
Due to their prominence in the literature, we  focus on two   active learners:

\subsubsection{Gaussian Process Models (GPM)} \label{ssec:gpm}

GPMs generate estimates by passing the available data through a range of kernels. In this way, they can
generate a  mean ($\mu$) and standard deviation ($\sigma$) for each prediction\cite{williams1995gaussian,brochu2010tutorial}. The   Upper Confidence Bound (UCB) acquisition function cab use $\mu,\sigma$ to guide the selection of the next
example to label. UCB recommends labeling the example $x$ that maximizes:

\begin{equation}
UCB(x) = \argmax_x (\mu(x) + \kappa \sigma(x))
\end{equation}
(where $\kappa$ is a constant).
Early in the reasoning, when little is known, the variances are large and the $\kappa \sigma(x)$ term dominates. Later,
as more data reduces the variance, 
UCB converges to just $\mu(x)$.
In this way, UCB adapts from
{\em exploring} regions of large variance
to {\em exploiting} regions with best mean prediction.

\subsubsection{Tree-structured Parzen Estimator (TPE)} \label{ssec:tpe}

The Tree-structured Parzen Estimator (TPE) differs from Gaussian Process methods by separately modeling the conditional distributions $p(x|y)$ and the marginal distribution $p(y)$ \cite{bergstra2011algorithms}. TPE splits the data based on objective values into two distinct groups, defined by a threshold $y^*$. Specifically, the conditional probability is represented as:

\begin{equation}
p(x|y) = \begin{cases}
l(x), & \text{if } y < y^* \\
g(x), & \text{if } y \geq y^*
\end{cases}
\end{equation}
Here, $l(x)$ represents the distribution of high-performing
observations, whereas $g(x)$ corresponds to the lower-performing
observations. The threshold $y^*$ is determined using a quantile
$\gamma$, ensuring $p(y < y^*) = \gamma$. The primary goal of TPE
is to select configurations that maximize the probability under
$l(x)$ and minimize it under $g(x)$, optimizing:
\begin{equation}
\arg\min_x \frac{g(x)}{l(x)}
\end{equation}


\subsubsection{Model Selection}

\here{R2c} \BLUE
In this study, our   objective is to compare Synthcore (an LLM-based synthesis method) against   symbolic baselines (UCB, TPE, etc), rather than to benchmark different LLMs against one another. Accordingly, the goal is not to identify which LLM performs best, but to ensure that the LLM used in the evaluation is sufficiently capable to provide a fair and meaningful comparison.

Hence we select Gemini 1.5 Pro, a state-of-the-art model at the time of the study, whose large context window (1M tokens) and strong factual reasoning abilities enable it to handle the rich metadata, multi-example prompts, and high-dimensional structures required for our task. 

\BLACK

(Aside: In the prior experiments from \cite{senthilkumar2024can}, both gpt-o1 and gemini-1.5-pro consistently produced similar results for all categories of the datasets (low, medium, \& high dimensional). Therefore,  rather than focusing more on the performance of different models, we focus on comparing LLMs collectively versus other baseline methods. )

For completeness, we  also looked into other models, but
 Gemma-3 and Llama3.1
 have two orders of magnitude fewer variables
than Gemini 1.5 Pro, and they performed poorly on most of the datasets\footnote{This study was conducted before the release of
Gemini 2.5 on March 21, 2025.}.
Also
, DeepSeek-R1 has achieved much recent publicity, but we found
we could not run it locally. Furthermore, the quotes we received for on-line experimentation of that model were prohibitively expensive\footnote{Our experiments require 20 repeats of a 10-way ensemble for 49 data sets. Total quotes
we received from that kind of inference ranged from \$5,000
to \$20,000. Our
lab supports
 a dozen graduate students,
each trying to write, at most, 3 papers per year. Supporting
this kind of inference would
hence cost up to 
$(3*12*20,000)=
\$720,000$, annually.}.
\begin{table}[!t]

\caption{Prompt template for few shot learning, here the two variables are 
{\tt meta} which is a table containing the meta data of the dataset
(name of column, min and max values,
mean or mode, standard deviation or entropy).
{\tt table} which are all the rows
of that dataset.}\label{tbl:fsl}

\centering\footnotesize

\begin{tabular}{|p{0.95\linewidth}|}
\hline

\rowcolor{blue!20} 
\textcolor{black!70}{\textbf{System Message:}} \\
\texttt{You are given a dataset with several features. The rows have been categorized into \textbf{"Best"} and \textbf{"Rest"} examples based on their overall performance. Below are the key features and their descriptions from the dataset:}\\
... \\
\textcolor{black!70}{\texttt{\{rows\_to\_markdown(meta)\}}} \\[1em]

\hline

\rowcolor{teal!15} 
\textcolor{teal!80!black}{\textbf{Human Message:}} \\
\texttt{Given Examples:}\\[1em]
\textcolor{black!70}{\texttt{\{rows\_to\_markdown(table)\}}} \\[1em]

\hline

\rowcolor{gray!15} 
\textcolor{black!80}{\textbf{Task:}} \\
\texttt{Generate an examples that is Better:} \\
\texttt{This should outperform the given \textbf{"Best"} examples by optimizing the relevant features to better combinations.} \\[1em]

\texttt{Consider the inter-dependencies between features, and ensure that the generated examples follow logical consistency within the dataset's context.} \\[1em]

\texttt{Return the output in the same markdown structure:} \\

\hline
\end{tabular}
\end{table}
\subsection{Prompting Strategies with LLMs}\label{prompt}
 In an attempt to emulate human reasoning,
we used LLMs to generate the warm starts needed for
our active learners. These warm starts are created
using the $N$-shot prompt tactic  of Table~\ref{tbl:fsl}.  In summary:
\begin{itemize}
\item The initial {\em System Message}
introduces the LLM to attribute names of a data set. 
\item The subsequent {\em Human Message}
then shows the LLM  $N$ examples, at random, and divide them into half ``best'' and ``rest'' (using Equation~1).
\item The {\em Task} prompt  asks the LLM  to synthesize an example better than all the `best'' and ``rest''  seen so far,   
\item Label and return the row nearest this better synthetic example.
\end{itemize}
We call this an $N$-shot learners where $N$ is the number of
labelled examples given as part of the prompt . For example,
if the {\em Human Message} shows  $N=6$ labelled rows, then we would call that a 6-shot learner.

SynthCore is a 
 \[L + M*N\]
shot  learner where $M$ is the number of
 repeated trials of our $N+1$-shot learners. Unlike Chain of Thought  or MAD   (Multi-Agent   Debates~\cite{smit2023we}), each $M$ trial has no knowledge of the results of the other trials. Rather:
 \begin{itemize}
 \item After labeling $L$  randomly selected rows,
 \item Repeat $M$ times
 \begin{itemize}
 \item Reset the context window of the LLM. 
 \item
 Apply the  $N$-shot learning of Table~\ref{tbl:fsl}.
 In that process, the LLM is given    all the column names and all the x values of the unlabeled rows.
 \end{itemize}
 \item All the $L + M*N$ labels generated by the this loop
 are then sorted via Equation~1.
 \item The best example in that sort is then returned. 
 \end{itemize}
 We recommend this approach for two reasons:
 \begin{enumerate}
 \item
Basic
 $N$-shot learner fails for higher-dimensional optimization
 problems. On the other hand,
 as shown in our {\em Results} section,
 SynthCore's $L+M*N$ approach performs very well,
 \item
 SynthCore is very simple to implement: it is just a wrapper around the established prompt strategy of   Table~\ref{tbl:fsl}.
 \end{enumerate}
Further to the second point,  since SynthCore is so simple,   it is worth asking
 if it merits publication at venues like this journal.
 
 The software engineering and AI literature has many examples where 
something can be very simple, yet still be very significant\footnote{Code reviews  is ``just'' asking that   at least one other developer examines and approves changes before they are merged. Rigby and Bird~\cite{rigby13} demonstrated that this simple addition to a DevOps pipeline yields substantial benefits to organizational knowledge, increasing programmers' understanding of the broader system by 60\% to 150\%. Furthermore, in the projects they studied,  this change transformed code reviews from a defect-finding task into a collaborative problem-solving activity.} \footnote{Boosting is ``just'' resampling technique where learner $i+1$ places increased focus on examples misclassified by learner $i$~\cite{FREUND1997119}. Its co-inventor, Yoav Freund, famously remarked:
{\em Boosting is the best 20-line shell script ever written}.
The AdaBoost script, while very short,  catalyzed a revolution in machine learning. With minimal logic, AdaBoost demonstrated how to transform weak learners into strong ones, laying the groundwork for modern ensemble methods such as XGBoost and LightGBM.}.
While a research {\em result} might be quite simple,
the research {\em effort} to achieve that result can be considerable.
SynthCore's  current   simplicity is the result of months of systematic exploration and overcoming obstacles:
\begin{itemize}
\item Our initial results from 
Table~\ref{tbl:fsl} where obtained using $N=4$ examples. Progressively increasing $N$  did not yield the desired performance gain. Furthermore, since the LLM had to reflect on all rows at each step of its {\em Task}, this approach became
excessively slow.
\item Chain-of-thought (CoT) prompting was investigated, which typically requires the LLM to generate step-by-step reasoning. However, developing an explanation module that allowed the LLM to offer genuinely insightful reflections on our specific tasks proved to be challenging. This experience aligns with findings from other researchers who note that tables can be particularly difficult for models primarily trained on textual sequences, and that complex reasoning over them remains a hurdle even with techniques like normalization~\cite{pourreza2023normtab}.
\item We also explored Multi-Agent Debates (MAD)~\cite{smit2023we}, but the computational costs associated with running multiple agents were prohibitive.
\begin{itemize}
\item
Our experiments were applied to 49 data sets.
\item
In each experiment, $M*N$ times, an LLM has to reflect
on the x-values on the unlabeled rows. As shown in Figure~\ref{dataset}, the MOOT problems can range up to 90,000 rows.
\item
To ensure statistical validity and account for the non-deterministic behavior of the LLM, all our   experiments were repeated 20 times using different random seeds.
\end{itemize}
\end{itemize}
In the end what lead us to SynthCore was the observation
was that through all the above, our methods did sometimes stumble on the optimal solution\footnote{Given data like Table~\ref{dataset}, a solution is ``optimal''  if it is the row closest to the ideal point, as judged by Equation~1.}.
This observation called to mind all the research on ensemble
methods   which lead us to try
SynthCore's  $L+M*N$ ensemble approach.

Before continuing, we make one technical point. After labeling $N$ items, the $N$-shot learner has to
label one more item (the best example seen so far).
Accordingly, the total number of labels required for these methods
is $L+M*(N+1)$.


\subsection{Experimental Rig}
\BLUE
\here{R4d} Our choices of the initial annotation budget $L$ and the later sampling budget $M$ follow directly from the Near Enough Optimization (NEO) framework. NEO adopts the view that (a) ``near enough is good enough,'' and (b) solutions within a small effect-size $\varepsilon$ of the optimum are indistinguishable. Under this assumption, sampling is not required to locate the global optimum, but only to encounter \emph{any} solution lying within an $\varepsilon$-neighbourhood of it.

If solutions are randomly shuffled in the dataset (Assumption~1) and values within $\varepsilon$ are indistinguishable (Assumption~2), then the probability of drawing an $\varepsilon$-optimal solution in a single trial is proportional to $\varepsilon$. The waiting time to the first success can therefore be modelled as a geometric process~\cite{ross2014introduction}. After $n$ draws, the probability of having observed at least one success is:
\[
1 - (1-\varepsilon)^n \ge C.
\]

Solving for $n$ yields:
\[
n(C,\varepsilon) = \frac{\log(1-C)}{\log(1-\varepsilon)}.
\]
Cohen's rule~\cite{cohen2016power} is that results are statistically distinguishable  by more than a small effect if they different by half a standard deviation. Assuming a normal distribution with the span
$\pm 3$ then  $\varepsilon=0.5/6$. 
Hence, to be 95\% sure that we have
found solutions that are 
 indistinguishable  from the best, we need the following number of samples:
\[
n(C=0.95) \approx 60
\]
The NEO analysis indicates that $\approx60$ samples are sufficient to reach an $\varepsilon$-near-optimal region. However, once the initial $L$ samples anchor the search, the remaining adaptive rounds need not individually reach the full theoretical requirement. Instead, they serve as incremental refinement steps.

We therefore select $M = 20$ \& $M=30$ as a deliberate engineering compromise:
\begin{itemize}
    \item $M=10$ was too small to provide meaningful refinement in most datasets.
    \item $M=40$ offered marginal improvements while increasing labeling costs.
    \item $M=20$ \& $M=30$ consistently achieved the best trade-off between performance and annotation expense, whereas $M=20$ performed better for low-dimensionality problems, while $M=30$ performed better in medium and high-dimensionality problems.
\end{itemize}

Using the nomenclature from the previous section, we say:
\begin{itemize}
\item $B=L+M*(N+1)$
\item $L=0.6\*B$
\item $M=20$ i.e., perform 20 repeats to test whether our results hold across a range of randomly selected initial conditions.
\end{itemize}

\BLACK

We ran all our active learners with budgets in the range of {20..100} for 20 repetitions to obtain the statistics for the Scott-Knott test. Budgets up to 100 were chosen since, as seen below, no performance improvements were observed over 50 samples.

\subsection{Statistical Methods (Scott-Knott)}\label{stats}

  We run Scott-Knott tests for all the datasets across different treatments, summarize the results in the above subsections, and report the percentage of times a treatment (e.g., "SynthCore") appears in rank $|n|$.

 Results were ranked  using a combination of the Scott-Knott,
 bootstrap, and Cliff's Delta procedure~\cite{SK74}. 
Scott-Knott recursively partitions sorted treatment means (derived from their performance metric, e.g., Chebyshev distances from the repeats) into statistically distinct, non-overlapping groups \cite{GB91}. Each partition maximizes the between-group sum of squares:
\begin{equation}
SS_B = n_1(\bar{x}_1 - \bar{x})^2 + n_2(\bar{x}_2 - \bar{x})^2
\label{eq:ssb_concise}
\end{equation}
where $n_1, n_2$ are sub-group sizes with means $\bar{x}_1, \bar{x}_2$, and $\bar{x}$ is the combined group mean. 
A key advantage of the Scott-Knott method is that it creates statistically distinct, non-overlapping groups of means. This simplifies interpretation, as there is no ambiguity about which group has what treatment.

Scott-Knott recursively splits treatments   if   a significance test and an effect size test report that     groups are (a) statistically distinguishable and (b) different by more than a small effect.
Cliff's Delta ($\delta$) \cite{Cli93} is a non-parametric effect size that quantifies the magnitude of the difference between treatment performance distributions. It measures the probability of one distribution being stochastically greater than another~\cite{Cli93}.

Bootstrap resampling \cite{Efr79} assesses the statistical significance of the observed differences. This non-parametric technique creates empirical sampling distributions by resampling with replacement from each treatment's 20 performance values. It tests if the observed deviations from random variation are significant, which is valuable for hypothesis testing and model validation, particularly with limited or non-normally distributed data \cite{ET93}.

\section{Results}\label{results}

Before conducting comparative evaluations of different methods, it is prudent to first document the baseline performance of our proposed approach. Why?  Reporting only relative results, without first establishing an absolute baseline,   undermines the significance of the results and should be avoided.
All too often, we read research papers where authors   overlook this step and then report seemingly impressive improvements—such as a X\% gain— which is
 misleading when the baseline itself is unacceptably low. 
 
 \begin{figure}
    \centering
    \includegraphics[width=0.75\linewidth]{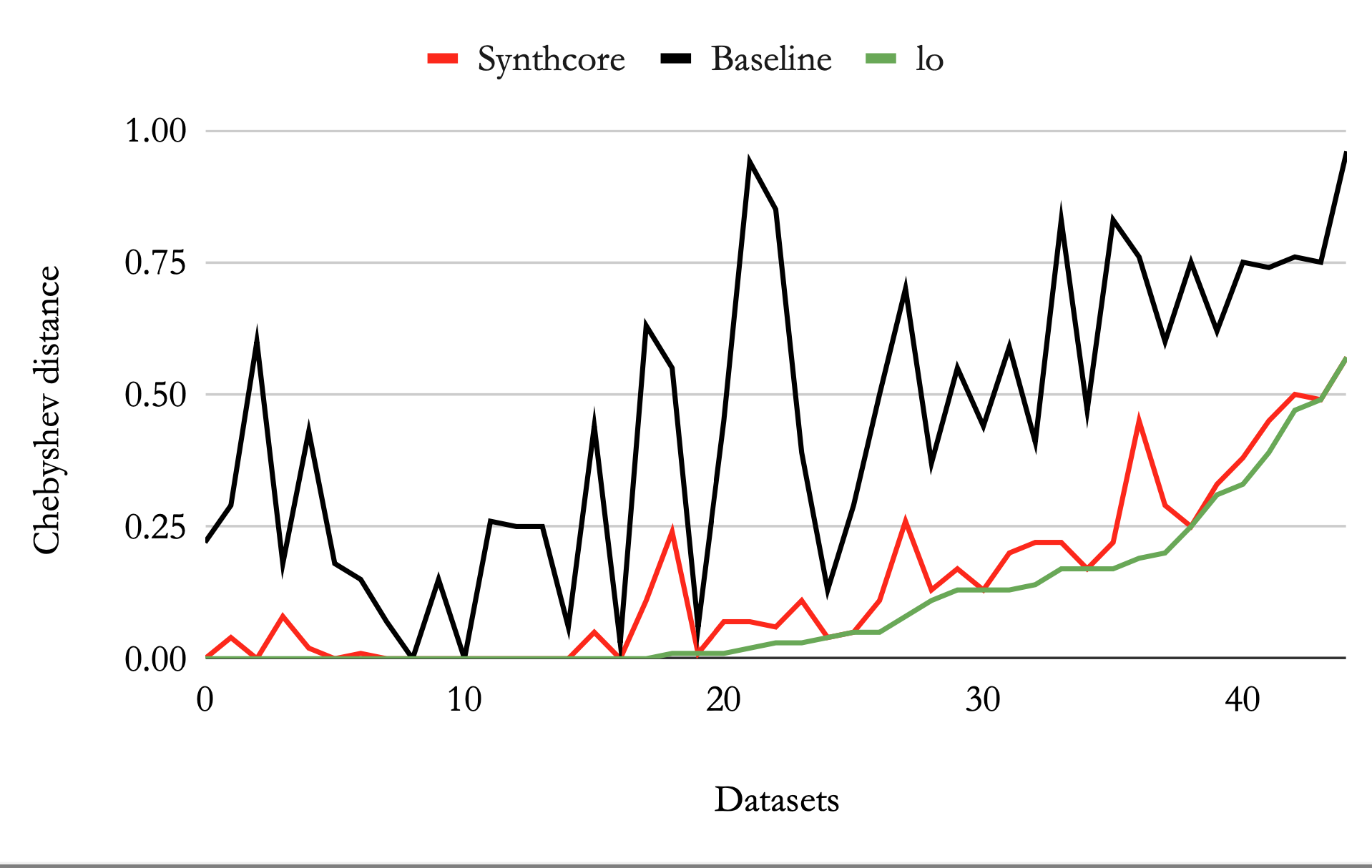}
    \caption{Performance of Synthcore 20, 
    (shown in red) fall very close to the
    best possible optimal (shown in green).}
    \label{lobaseline}
\end{figure}

 Figure~\ref{lobaseline} shows
 mean results from twenty runs of Synthcore (budget $B=20$ labels, LLM‐generated warm start):
\bi
\item
The 
green curve sorts data by their best row that is closest to “heaven” (see Equation \ref{eq:ch}).
\item
The black curve represents the untreated ``before'' condition of our data. This curve shows the  average distance to heaven if all rows.
If we just picked random
rows, we would usually achieve the results shown in black.
\item 
The red curve shows the rows
found by Synthcore.
\ei
By our design, every optimization falls between the black (mean) and green (minimum) lines.
Hence, the closer
the red curve falls to the green curve, the better the performance. The red curve’s proximity to green shows that Synthcore delivers strong results.

(Aside: not shown in Figure~\ref{lobaseline} are
statistical tests that rank Synthcore's performance against other   state-of-the-art methods. For those statistical comparisons, see below.)

\begin{figure}
    \centering
    \includegraphics[width=0.8\linewidth]{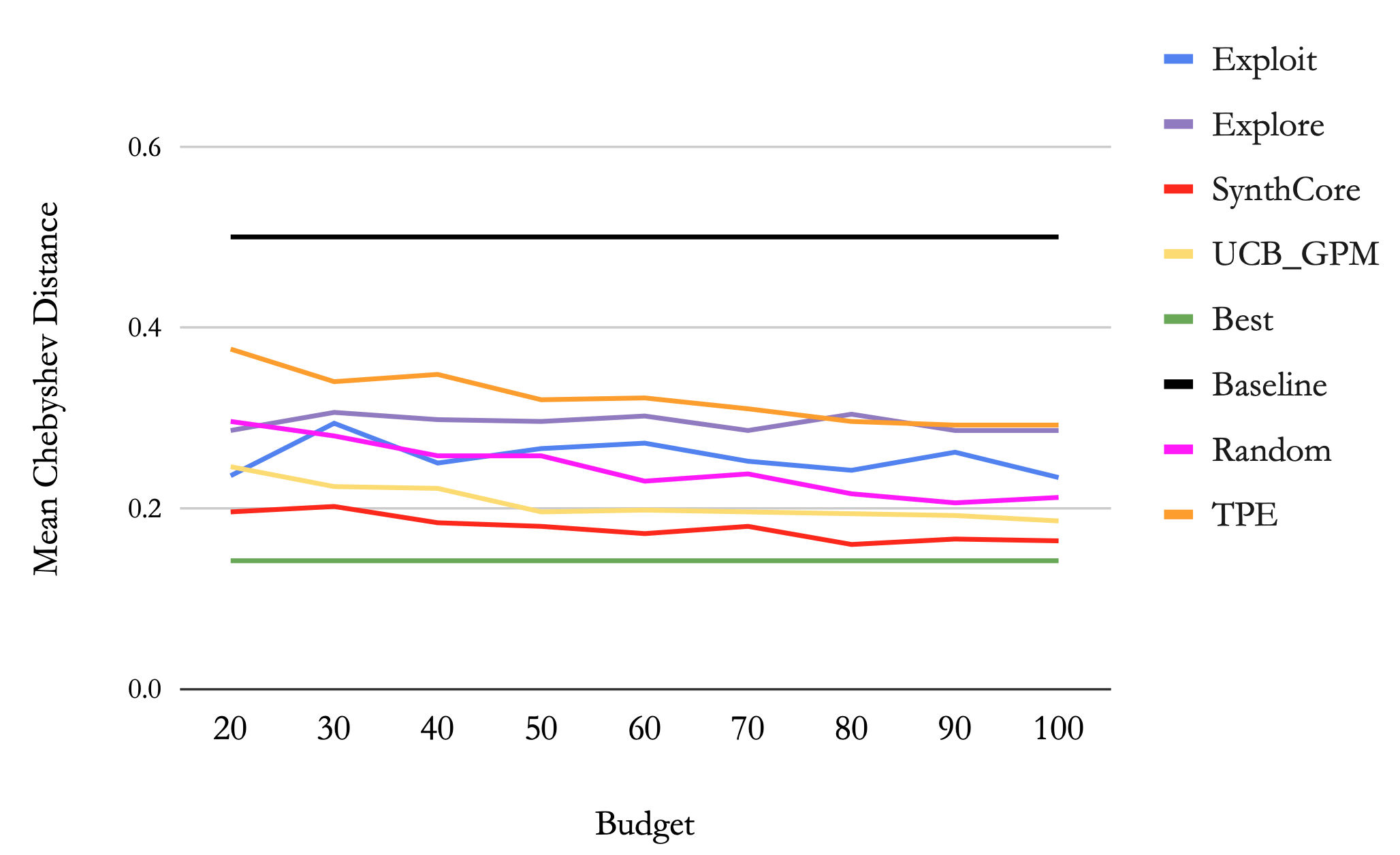}
    \caption{Performance of different active learners w.r.to budgets. Best is the average heaven values for all the datasets}
    \label{budget-vs-heaven}
\end{figure}

\begin{figure}
    \centering
    \includegraphics[width=0.8\linewidth]{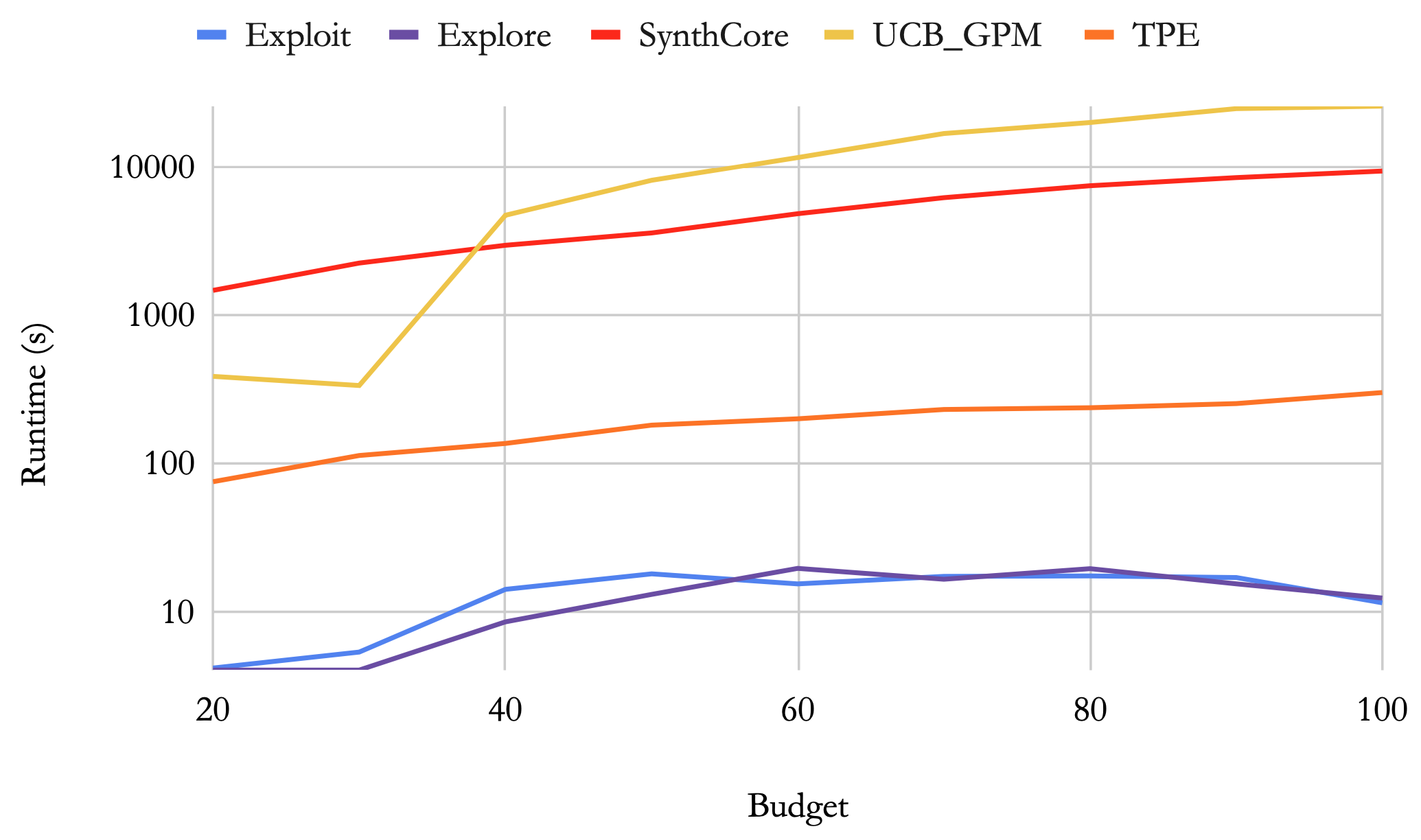}
    \caption{Average Run times of different treatments, y axis is time in seconds scaled to logarithmic values and x axis is budget ranging from 20 to 100}
    \label{runtimes}
\end{figure}

Figure~\ref{lobaseline} showed   Synthcore's benefits. Those benefits came at some cost:
\bi
\item
The cost of collecting labels;
\item
The CPU runtime cost.
\ei
  Recall that the Figure~\ref{lobaseline}   achieved at a cost of just a few labels ($B=20$). 
  Using 
   Figure~\ref{budget-vs-heaven} and Figure~\ref{runtimes}
 we can now justify that budget:
 \bi
 \item
 Figure~\ref{budget-vs-heaven}
 shows mean distance to heaven across all data sets in twenty repeated trials of our different optimizers.
  The bottom curve shows  the mean distance of heaven of the best row in each data set. 
 The x-axis shows how performance changes as the budget increases. The
 statistical tests of \S\ref{stats}
 shows no statistical different between the $B=20,30,50,100$ results.
\item
Figure~\ref{runtimes} shows the runtimes associated with budget sizes associated with varying budgets.
Comparing budgets of 20 and 100,
we note that Synthcore is much slower for larger budgets; e.g.
at $B=1000$ it is five times slower than at $B=20$.
 \ei
In summary, the (a)~runtimes are faster at smaller budgets;
and (b)~there is no 
 statistically significant optimization improvement at larger budgets. Accordingly,
 the rest of out results will be reported using $B=20$.

 Before continuing, we answer one frequently asked question: Why the orders of difference in run times between the lower and upper curved of Figure~\ref{runtimes}?
 The two lower curves from simple Bayesian methods that can update their models for every row,
 in linear time. The upper 
 two curves come from methods that are far more reflective. Specifically,
 the upper methods made decision by   (a)~reflecting over  a large space of possible models (UCB\_GPM),
 or (b)~reflecting over a large context window that struggled to pay
  attention to hundreds to thousands of rows (SythnCore). Hence it is hardly surprising
 that that more reflective methods (UCB\_GPM and SynthCore) are far slower than the other methods.

\subsection{Answers to Research Questions}

Have documented the absolute performance
of SynthCore, we now move to statistical
comparisons between different methods.

These comparisons are shown in Tables~\ref{lo0}, \ref{med0}, \ref{hi0} and 
Tables~\ref{lo}, \ref{med}, \ref{hi}
Each cell in this table shows the percentage of times (in 20 repeats) that a particular treatment was ranked best (i.e.  rank=0), or some other rank, by the statistical methods
of \S\ref{stats}. The most important part of these tables are the rank=0 results. For example, top left of
 Table~\ref{lo0}, we see that 100\% of the time, an LLM method achieved a top rank.

 The red results in Tables~\ref{lo0}, \ref{med0}, \ref{hi0} come from prior work 
 by
 Senthilkumar et al.~\cite{senthilkumar2024can}
 that ran the few-shot procedure of
 Table~\ref{tbl:fsl}, only once (no ensembles).
 Here we see the negative prior result that inspired this  study:
 \bi
 \item
 Tables~\ref{lo0}, \ref{med0} that LLM warm start method did as well, or better than anything else.
 \item
  But in Table~\ref{hi0}, LLMs were only half as good as the best method (UCB\_GPM).
 \ei
 The blue results in 
 Tables~\ref{lo}, \ref{med}, \ref{hi}
 show the performance improvements achieved by SynthCore's
 ensemble methods. Across all our data,
 SynthCore performed better than everything
 else. Hence we say:

{\textbf{RQ1:} {\em{Is ensemble few shot learning better than standard few shot learning for SE Active Learning?}}  
Answer: {\bf Yes}.

\begin{adjustbox}{scale=1}

 \begin{minipage}[t]{0.5\textwidth}

\begin{table}[H]
\centering
\scriptsize

{\bf Column 1 \\ Prior results, from~\cite{senthilkumar2024can}. \\  (no ensembles).}
\vspace{5mm}

\begin{tabular}{c|c@{~}c@{~}c@{~}c@{~}c@{~}c@{~}c} 
&\multicolumn{7}{c}{\textbf{Scott-Knott Rankings}}\\

  & 0 & 1 & 2 & 3 & 4 & 5 & 6 \\
\hline
LLM Warms  & \colorbox{darkred}{\makebox[0.4cm][l]{\textbf{100}}} &  & &  &  &  &  \\
Exploit  & \colorbox{red}{\makebox[0.3cm][l]{73}} & 18 & 9 & & & &  \\
UCB\_GPM & \colorbox{red}{\makebox[0.3cm][l]{55}} & \colorbox{lightred}{\makebox[0.3cm][l]{27}} &  & 9& &   \\
TPE & \colorbox{lightred}{\makebox[0.3cm][l]{27}} & \colorbox{lightred}{\makebox[0.3cm][l]{27}} & & 9 & 18 & 9\\
Explore   &   9 & 18 & 18 & \colorbox{lightred}{\makebox[0.3cm][l]{27}}& & &  \\
random    & \colorbox{red}{\makebox[0.3cm][l]{36}} &  & 9 & 9 & 9 & 9 \\
Baseline  & & & & & 9 &\colorbox{lightred}{\makebox[0.3cm][l]{18}} &  \\
\hline
\end{tabular}
\caption{Low dimensional   ($x < 6$).}\label{lo0}
\end{table}


\begin{table}[H]
\centering
\scriptsize
\begin{tabular}{c|c@{~}c@{~}c@{~}c@{~}c@{~}c@{~}c}  
&\multicolumn{7}{c}{\textbf{Scott-Knott Rankings}}\\

  & 0 & 1 & 2 & 3 & 4 & 5 & 6 \\
\hline

UCB\_GPM & \colorbox{darkred}{\makebox[0.3cm][l]{\textbf{64}}} & 18 & 9 & 9 &  & &9 \\
LLM Warms  & \colorbox{darkred}{\makebox[0.3cm][l]{\textbf{64}}} &  9& 18&  & &  &  \\
Exploit  & \colorbox{red}{\makebox[0.3cm][l]{45}} & 18 & \colorbox{lightred}{\makebox[0.3cm][l]{27}} & 18 &  10 & &  \\
TPE & \colorbox{red}{\makebox[0.3cm][l]{36}} & 18 & 18 & 18 & & 9\\
random    & \colorbox{lightred}{\makebox[0.3cm][l]{27}} & 18 &  \colorbox{lightred}{\makebox[0.3cm][l]{27}}& 18 & &  \\
Explore   &   27 & 9 & 18 & \colorbox{lightred}{\makebox[0.3cm][l]{20}} & & 10& 10 \\
Baseline  & 9 & 9& 9 &  &  & 18 & \\
\hline
\end{tabular}
\caption{Medium dim.   ($5 < x < 11$).}\label{med0}
\end{table}

\begin{table}[H]
\centering
\scriptsize
\begin{tabular}{c|c@{~}c@{~}c@{~}c@{~}c@{~}c@{~}c} 
&\multicolumn{7}{c}{\textbf{Scott-Knott Rankings}}\\
  & 0 & 1 & 2 & 3 & 4 & 5 & 6 \\
\hline

UCB\_GPM & \colorbox{darkred}{\makebox[0.4cm][l]{\textbf{100}}} &  & & &  & &  \\
LLM Warms & \colorbox{red}{\makebox[0.3cm][l]{\textbf{53}}} & 13 & 7& 13 &  &  &  \\
Exploit  & \colorbox{red}{\makebox[0.3cm][l]{\textbf{47}}} & 13 & 13 & 13 &   & &  \\
Explore   & 7 &  & 13 & 20 & 20& &  \\
TPE & \colorbox{red}{\makebox[0.3cm][l]{\textbf{40}}} & 13 & 7 & 13 & 7 & 7 \\

random    & &\colorbox{lightred}{\makebox[0.3cm][l]{27}} & \colorbox{lightred}{\makebox[0.3cm][l]{27}} & 20& 13 & 13 \\
Baseline & &7 & 13&  & 9 & 27 &  \\
\hline
\end{tabular}
\caption{High dimensional   ($x > 10$).}\label{hi0}
\end{table}

\end{minipage}

\hspace{0.01\textwidth} 
\begin{minipage}[t]{0.5\textwidth}
\begin{table}[H]
\centering
\scriptsize
{\bf Column 2 \\ Our results \\  (with few-shot ensembles).}
\vspace{5mm}

\begin{tabular}{c|c@{~}c@{~}c@{~}c@{~}c@{~}c@{~}c} 
&\multicolumn{7}{c}{\textbf{Scott-Knott Rankings}}\\

  & 0 & 1 & 2 & 3 & 4 & 5 & 6 \\
\hline
SynthCore  & \colorbox{darkblue}{\makebox[0.3cm][l]{\textbf{86}}} & 7  & &  &  &  &  \\
Exploit  & \colorbox{blue}{\makebox[0.3cm][l]{\textbf{71}}} & 7 & 7 & 7 &  7 & &  \\
UCB\_GPM & \colorbox{blue}{\makebox[0.3cm][l]{\textbf{43}}} & \colorbox{lightblue}{\makebox[0.3cm][l]{29}} & 7 & 7& 14 & &  \\
TPE & 7 & \colorbox{lightblue}{\makebox[0.3cm][l]{29}} & 7 &  14&  14 \\
Explore   &   7 & 7 & \colorbox{lightblue}{\makebox[0.3cm][l]{29}} & \colorbox{blue}{\makebox[0.3cm][l]{36}} & 14& 7& 7 \\
random    & 7 & \colorbox{blue}{\makebox[0.3cm][l]{36}} & \colorbox{lightblue}{\makebox[0.3cm][l]{29}} & 14 & 7 & 7 \\
Baseline  & & & & & 9 & \colorbox{lightblue}{\makebox[0.3cm][l]{21}} &  \\
\hline
\end{tabular}
\caption{Low dimensional   ($x < 6$).}\label{lo}
\end{table}
 \begin{table}[H]
\centering
\scriptsize
\begin{tabular}{c|c@{~}c@{~}c@{~}c@{~}c@{~}c@{~}c} 
&\multicolumn{7}{c}{\textbf{Scott-Knott Rankings}}\\

  & 0 & 1 & 2 & 3 & 4 & 5 & 6 \\
\hline
SynthCore  & \colorbox{darkblue}{\makebox[0.3cm][l]{\textbf{85}}} &  & &  &  8&  &  \\
UCB\_GPM & \colorbox{blue}{\makebox[0.3cm][l]{\textbf{77}}} & 8 & 8 & &  & & \\
Exploit  & \colorbox{blue}{\makebox[0.3cm][l]{\textbf{69}}} &  & 8 & 8 &  8 & &  \\
TPE & \colorbox{blue}{\makebox[0.3cm][l]{39}} & 8 & 8 & 15& 15 & \\
random    & \colorbox{blue}{\makebox[0.3cm][l]{31}} & 15 &  \colorbox{blue}{\makebox[0.3cm][l]{31}}& 15 & &  \\
Explore   &   9 & 9 & \colorbox{blue}{\makebox[0.3cm][l]{41}} & \colorbox{darkblue}{\makebox[0.3cm][l]{23}} & & 8& 8 \\
Baseline  & 8& & 15& 8& 15 & 15 &  8\\
\hline
\end{tabular}
\caption{Medium dim.   ($5 < x < 11$).}\label{med}
\end{table}

\begin{table}[H]
\centering
\scriptsize
\begin{tabular}{c|c@{~}c@{~}c@{~}c@{~}c@{~}c@{~}c}
&\multicolumn{7}{c}{\textbf{Scott-Knott Rankings}}\\

  & 0 & 1 & 2 & 3 & 4 & 5 & 6 \\
\hline
SynthCore  & \colorbox{darkblue}{\makebox[0.3cm][l]{\textbf{71}}} & 5 & 11&  8&  &  &  \\
UCB\_GPM & \colorbox{darkblue}{\makebox[0.3cm][l]{\textbf{68}}} & 11 & 11 & &  & & 5 \\
Explore   & \colorbox{blue}{\makebox[0.3cm][l]{37}} & 11 & \colorbox{lightblue}{\makebox[0.3cm][l]{21}} & 5 & 16& &  \\
TPE &\colorbox{blue}{\makebox[0.3cm][l]{32}} & 16 &16 & 5 & 5 & 5 \\
Exploit  & 16 & 16 & \colorbox{blue}
{\makebox[0.3cm][l]{32}} & 11 &  5 & 5& 5 \\
random    & &\colorbox{lightblue}{\makebox[0.3cm][l]{21}} & \colorbox{lightblue}{\makebox[0.3cm][l]{21}} &  16& 11 & 11 \\
Baseline  & & & 5& 11& 11 & \colorbox{lightblue}{\makebox[0.3cm][l]{26}} & 5 \\
\hline
\end{tabular}
\caption{High dimensional   ($x > 10$).}\label{hi}
\end{table}

\vspace{5mm} 
\end{minipage}

\end{adjustbox}

\newpage
Moving on to our second research question,
this work was motivated by a prior study that reported
    LLMs failed   on
    higher-dimensional. Therefore,
    in our results, we must pay particular
    attention to this kind of data.

    {\textbf{RQ2:} {\em {Does this improvement hold for high dimensional data?}}  Comparing Table~\ref{hi0} and
    Table~\ref{hi}, we see, for high-dimensional,
      ensembles thrive where solo-based
    reasoning fails. Hence for this question  we answer {\bf Yes}.

 To say all this another way,
 for this kind of optimization, LLMs need to be encouraged to divulge their knowledge. 
 A single prompt is not enough, nor is simple few shot learning. Rather, it is needed to run multiple sessions with an LLM, then aggregate the results.

\section{Discussion}

SynthCore demonstrates that large language models, though 
unreliable as single annotators, can become powerful contributors 
when treated as ensembles of weak, diverse learners. While past 
studies have shown that few-shot prompts struggle in 
high-dimensional SE tasks, our results indicate that diversity in 
prompting can significantly mitigate this limitation. Across 49 
datasets, SynthCore achieved consistent gains over traditional 
optimization baselines and even surpassed adaptive Bayesian 
learners in many cases.

One reason SynthCore works is that ensemble prompting reduces 
overfitting to prompt phrasing or context. Each prompt acts as a 
sample from the LLM’s latent capability space, and their 
aggregation offers resilience against individual prompt 
failures. This is especially valuable in SE optimization, where 
initial warm-starts are crucial to guiding the search process. 
Rather than placing all trust in a single LLM prediction, 
SynthCore leverages variation to increase coverage of plausible 
solutions early in the learning loop.

However, the benefit of ensembling is not uniform across all tasks.
Our experiments show that ensemble gains are greatest in 
moderate-dimensional spaces with clear trade-offs, such as 
project planning or Makefile tuning. In simpler tasks, single-shot 
learners often suffice. Few-shot prompting typically shows the 
highest returns when moving from 0 to 1 shot, with diminishing 
returns after 4 shots. Ensembles counteract this drop-off, 
particularly in higher-dimensional settings.

Crucially, SynthCore avoids complex multi-agent orchestration. 
Unlike chain-of-thought prompting or debate-style sampling, it 
does not rely on model introspection or self-consistency checking.
Instead, it uses independent prompts with no cross-talk. This 
simplicity makes SynthCore faster to run, easier to adapt to new 
domains, and more robust to failures in reasoning depth or 
hallucinated explanations. In practice, this makes SynthCore more 
applicable in real-world SE pipelines where latency and 
predictability matter.

\here{R3d}
\BLUE
For practitioners, we recommend using SynthCore following the workflow shown in Figure~\ref{recommend}. Our results indicate that SynthCore offers the largest performance gains in complex, high-dimensional SE optimization tasks where human heuristics are difficult to encode. For SE optimization problems, we advise scaling the evaluation budget with the complexity of the data—using the number of independent variables as a practical indicator—and increasing SynthCore’s budget accordingly. However, SynthCore is not universally optimal. When strong SME expertise is readily available, relying on a human SME may be more cost-effective than invoking LLM ensembles. Likewise, for highly structured and common tasks such as sentiment analysis or bug detection, standard LLM prompting already performs well, and the additional API cost and runtime of ensemble prompting may not be justified.

\begin{figure}
    \centering
    \includegraphics[width=\linewidth]{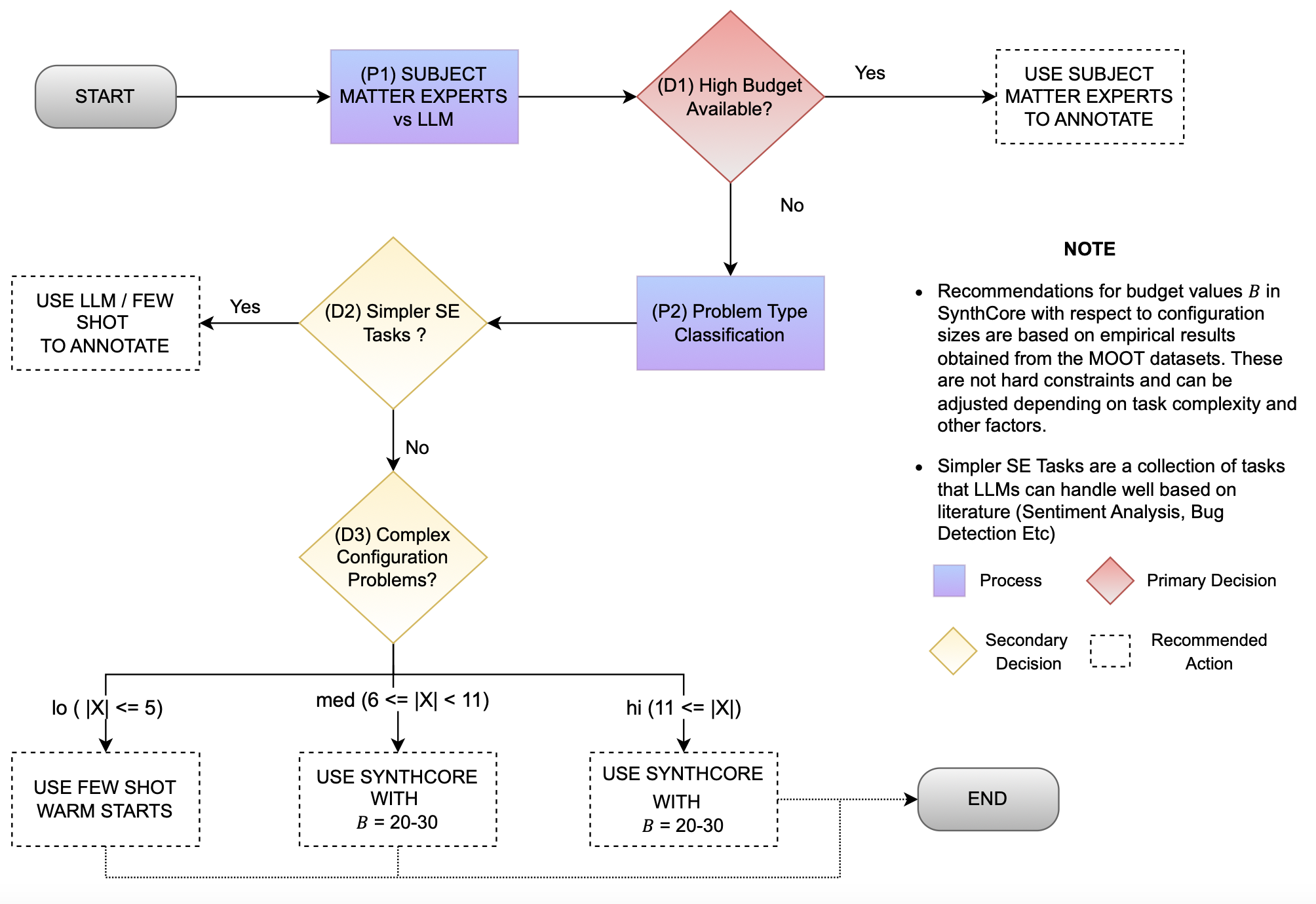}
    \caption{\protect\here{R3d-2} ~Recommendations to practitioners.}
    \label{recommend}
\end{figure}

\BLACK

SynthCore is also efficient. Compared to multi-stage LLM pipelines 
that require dialog, argument tracking, or self-refinement, 
SynthCore relies on multiple single-prompt calls—each simple, 
stateless, and fast. This architecture is well-suited to 
deployment within CI/CD workflows or as part of configuration 
tuning tools. Our results show that ensembles of size 3–5 provide 
most of the performance gains, making this approach viable even 
under strict budget constraints.

Beyond performance, SynthCore also surfaces a deeper point about 
LLM limitations. Prior work (e.g., Ahmed et al. at MSR 2025) has 
shown that LLM agreement with human annotators is not always a 
good proxy for truth. Our work shifts the question: rather than 
ask whether an LLM is “correct,” we ask whether a group of diverse 
LLM prompts can generate a fertile search space. Seen in this way, 
LLMs are not oracles—they are generative seeds for optimization.

\BLUE

\here{R3g}  More generally, we say that SynthCore functions as a subject matter expert amplifier, not a replacement. Its primary role in the pipeline is to compress the human expert's review surface. Instead of requiring an SME to laboriously generate and evaluate a large set of candidates---many of which are suboptimal---SynthCore filters and surfaces a small, high-quality candidate set for validation. This fundamentally shifts the SME's task from slow, exhaustive \emph{generation} to rapid, high-leverage \emph{validation}.

This transformation provides concrete benefits for efficiency and iteration speed. Without SynthCore, an SME might label over 100 examples to identify the 5 truly optimal configurations. With SynthCore, the SME's effort is focused on reviewing approximately 20 top candidates, validating, or making minor adjustments. This reduces the SME's  cognitive load significantly, as evaluating five promising designs is an intrinsically different task from shifting through fifty unknowns. Furthermore, human judgment is thus applied precisely where it is most needed: assessing ambiguous trade-offs, applying domain constraints the LLM may not possess, and integrating specific organizational context.

Regarding the need for human SME input, SynthCore reduces the required volume of input for a given quality threshold. It facilitates the democratization of the initial design phase. Junior engineers, leveraging SynthCore, can rapidly produce reasonable first drafts and solution proposals that previously required the intervention of a senior SME. Senior SMEs can then dedicate their limited, high-value time not to initial exploration, but to final refinement and tackling complex, boundary-pushing engineering challenges.

\here{R3d-b} However, we do not claim SynthCore to be a universal solution for annotation in multi-objective optimization problems. While text and code are indeed high-dimensional modalities, modern LLMs handle them remarkably well due to their training objectives and architectural design. This is also reflected in the tasks explored by Ahmed et al., which were predominantly text- and code-centric (e.g., sentiment classification), where LLMs naturally excel.
In contrast, our focus is on tabular multi-objective optimization problems, a setting where LLMs typically struggle. Tabular data lacks the sequential structure and semantic regularities that LLMs exploit in language or code, making annotation and performance estimation significantly more challenging. For this reason, SynthCore is specifically designed for such structured, non-textual domains.

At the same time, we acknowledge that SynthCore may not generalize well to multimodal, ultra–high-dimensional domains such as audio or images, which fall outside the native capabilities of current LLMs. Addressing these more complex modalities would likely require future extensions that incorporate multimodal foundation models \cite{wang2023visionllm} capable of reasoning jointly over text, vision, and other signals.

\BLACK

\subsection{Future Work}

We plan to explore transferability across iterations. Currently, 
SynthCore treats each ensemble as independent, but a sequential 
execution model—where insights from early prompts inform later 
ones—could reduce sampling cost while improving guidance. This 
strategy may enable higher efficiency with fewer prompt calls.

A natural next application is Next-Release Planning (NACP), where 
trade-offs among value, effort, and delivery constraints align 
with SynthCore's strengths in navigating sparse, 
multi-objective spaces. Here, ensembles could support decisions 
under uncertainty with little labeled data.

Another direction is extending SynthCore to structured domains 
such as patch generation, bug triage, or Makefile tuning. These 
require prompts that account for semantics and ordering. Ensemble 
diversity may help span different reasoning paths over these 
structures.

We also intend to benchmark SynthCore against expert-curated 
annotations—not just to evaluate predictive accuracy, but to 
assess alignment with human judgment and trust. This could 
clarify the role of LLMs as surrogates or collaborators.

Explainability remains a key open challenge. While SynthCore 
improves sampling, it does not make LLM reasoning more transparent.
Post-hoc interpretation of prompt choices or traceable 
rationales could make the system more suitable for regulated or 
high-stakes domains.

Finally, we plan to automate prompt selection using clustering or 
meta-learning. Rather than relying on static prompt sets, 
SynthCore could dynamically adjust ensemble makeup based on task 
feedback or historical performance, further improving generality 
and sample efficiency.

\subsection{Threats to Validity}

\paragraph{Construct Validity.} While our evaluations use standard 
multi-objective metrics, these may not capture qualities important 
to practitioners, such as long-term maintainability or developer 
trust. Further studies could compare SynthCore-generated configs 
against expert-labeled benchmarks or measure downstream effects on 
developer workflow.

\paragraph{Internal Validity.} Though care was taken to verify our 
implementations, there may be latent bugs or assumptions in our 
active learning and acquisition functions. Hyperparameters for 
GPM, TPE, and LLM prompting were chosen empirically, and small 
changes may affect outcomes. We encourage replication and 
sensitivity testing using our reproduction package.

\paragraph{External Validity.} Our 49 datasets cover a broad range 
of SE optimization problems, but not all real-world tasks. Tasks 
with strong semantic constraints—like patch generation or 
prioritization based on security context—may demand additional 
domain knowledge or hybrid LLM-human annotation schemes. Also, 
while we used open-source models, performance may differ for 
closed commercial APIs.

\section{Conclusion}

This work presents SynthCore, a simple but powerful ensemble 
prompting technique for LLM-based annotation and optimization in 
software engineering. By combining multiple few-shot prompts with 
no coordination or chain-of-thought overhead, SynthCore achieves 
state-of-the-art performance on high-dimensional SE optimization 
tasks—outperforming Bayesian methods, active learners, and 
single-prompt baselines. All results were achieved using only 
LLM-generated data, without human intervention.

Our findings reframe how LLMs should be used in SE. Rather than 
expecting correctness from a single prompt, we should treat LLMs 
as stochastic samplers of reasoning fragments. When these samples 
are aggregated via SynthCore, they yield diverse and effective 
search spaces for optimization—even on tasks where single prompts 
fail.

SynthCore’s value lies in its generality. It requires no 
fine-tuning, no deep prompt engineering, and no supervision. It is 
easy to implement, extensible to many task types, and efficient 
enough to use in budget-constrained settings. This makes it a 
practical building block for next-generation SE tooling.

\section*{Declarations}

\begin{description}
\item[\bf Source of Funding: ] \hfill \\
No funding bodies were involved in the creation
of this work.
\item[\bf Financial or non-financial interests: ] \hfill \\
The authors have no relevant financial or non-financial interests to disclose.
All authors certify that they have no affiliations with or involvement in any organization or entity with any financial interest or non-financial interest in the subject matter or materials discussed in this manuscript.

\item[\bf Ethical approval:] \hfill \\
Lacking human or animal subjects,
an ethical review was not required. 
\item[\bf Informed consent:] \hfill \\
This study had no human subjects.
\item[\bf Author Contributions:] \hfill \\
All this paper's experimental work was conducted by Lohith Senthilkumar.
Lohith Senthilkumar and
Tim Menzies contributed equally
to the writing of this paper. 
\item[\bf Data Availability Statement:] \hfill \\
To repeat and/or refine and/or refute this work, see our scripts and data at
\url{https://github.com/lohithsowmiyan/lazy-llm/tree/clusters}.
\item[\bf Conflict of Interest:] \hfill \\ The authors declared that they have no conflict of interest.
The authors have no competing interests to declare that are relevant to the content of this article.
\item[\bf Clinical Trial Number:] \hfill \\
Clinical trial number: not applicable.
\end{description}

\begin{acknowledgements}
If you'd like to thank anyone, place your comments here
and remove the percent signs.
\end{acknowledgements}

Authors must disclose all relationships or interests that 
could have direct or potential influence or impart bias on 
the work: 

\section*{Conflict of interest}

The authors declare that they have no conflict of interest.


\bibliographystyle{plain} 
\bibliography{123} 

\end{document}